\RequirePackage{silence}
\documentclass[aps,pra,twocolumn,showpacs,amsmath,amssymb,preprintnumbers,superscriptaddress,10pt]{revtex4-2}

\usepackage{graphicx}
\usepackage{float}
\usepackage{subfigure}
\usepackage{dcolumn}
\usepackage{bm}
\usepackage{hyperref}
\usepackage{siunitx}
\usepackage{amsmath,amssymb}
\usepackage{marvosym}
\usepackage{braket}
\usepackage{booktabs}
\usepackage{appendix}
\hypersetup{colorlinks=true, citecolor=blue, urlcolor=blue, linkcolor=blue}

\begin{document}
	
\title{Spectral side channels of wavelength-division multiplexer in quantum key distribution under laser damage}

\author{Binwu~Gao}
\thanks{These authors contributed equally}
\affiliation{College of Computer Science and Technology, National University of Defense Technology, Changsha 410073, People's Republic of China}

\author{Junxuan~Liu}
\thanks{These authors contributed equally}
\affiliation{College of Computer Science and Technology, National University of Defense Technology, Changsha 410073, People's Republic of China}

\author{Ekaterina~Borisova}
\affiliation{Russian Quantum Center, Skolkovo, Moscow 121205, Russia}
\affiliation{NTI Center for Quantum Communications, National University of Science and Technology MISiS, Moscow 119049, Russia}

\author{Hao~Tan}
\affiliation{China Telecom Quantum Information Technology Group Co., Ltd., Hefei 230088, China}

\author{Mingyang~Zhong}
\affiliation{College of Computer Science and Technology, National University of Defense Technology, Changsha 410073, People's Republic of China}

\author{Zihao~Chen}
\affiliation{College of Computer Science and Technology, National University of Defense Technology, Changsha 410073, People's Republic of China}

\author{Qingquan~Peng}
\affiliation{College of Electronic Science and Technology, National University of Defense Technology, Changsha 410073, People’s Republic of China}
\affiliation{College of Computer Science and Technology, National University of Defense Technology, Changsha 410073, People's Republic of China}

\author{Weixu~Shi}
\affiliation{College of Computer Science and Technology, National University of Defense Technology, Changsha 410073, People's Republic of China}

\author{Anastasiya~Ponosova}
\affiliation{Russian Quantum Center, Skolkovo, Moscow 121205, Russia}
\affiliation{NTI Center for Quantum Communications, National University of Science and Technology MISiS, Moscow 119049, Russia}

\author{Vadim~Makarov}
\affiliation{Russian Quantum Center, Skolkovo, Moscow 121205, Russia}
\affiliation{NTI Center for Quantum Communications, National University of Science and Technology MISiS, Moscow 119049, Russia}
\affiliation{Vigo Quantum Communication Center, University of Vigo, Vigo E-36310, Spain}

\author{Anqi~Huang}
\email{angelhuang.hn@gmail.com}
\affiliation{College of Electronic Science and Technology, National University of Defense Technology, Changsha 410073, People’s Republic of China}
\affiliation{College of Computer Science and Technology, National University of Defense Technology, Changsha 410073, People's Republic of China}

\date{\today}

\begin{abstract}
In the transmitter of a quantum key distribution (QKD) system, a wavelength-division multiplexer (WDM) is typically used to combine quantum and synchronization signals and is directly connected to the quantum channel. As a result, it becomes the first optical component exposed to laser-injection attacks. Therefore, understanding the behavior of WDMs under such attacks is essential for assessing the practical security of QKD systems. In this work, we systematically investigate the characteristics of WDMs under high-power laser illumination. Our experimental results show that certain WDM samples exhibit pronounced changes in their spectral features once the injected laser power surpasses a specific threshold. Taking the Trojan-horse attack as an illustrative example, we further perform a theoretical analysis of the resulting spectral side channel and show that it can reduce the maximum secure transmission distance to below $66.9\%$ of its original value. By combining experimental observations with theoretical modeling, this study advances the understanding of the influence of WDMs on the practical security of QKD systems.
\end{abstract}

\maketitle

\section{Introduction} \label{sec:Introduction}

        
Quantum key distribution (QKD) offers information-theoretic security for sharing symmetric keys between two communicating parties, Alice and Bob. Its security stems from the fundamental principles of quantum physics and is therefore immune to advances in computational power~\cite{bennett1984,RevModPhys.74.145,RevModPhys.81.1301,RevModPhys.92.025002}. However, the physical devices used in practical QKD implementations inevitably deviate from their idealized models, giving rise to security vulnerabilities~\cite{PhysRevA.74.022313,Lydersen2010,Lydersen:10,Xu_2010,PhysRevA.84.062308,Wiechers_2011,Gerhardt2011,PhysRevLett.107.110501,PhysRevLett.112.070503,PhysRevA.92.022304,PhysRevA.91.062301,7571108,PhysRevA.98.012330,PhysRevApplied.10.064062,PhysRevApplied.12.064043,PhysRevApplied.13.034017,e24020260,Chaiwongkhot2022,PhysRevA.106.033713,PRXQuantum.3.040307,PhysRevApplied.19.014048,PhysRevApplied.21.014026,Peng2025}. These device imperfections constitute side channels that may enable an eavesdropper (Eve) to gain information about the secret key.

Considerable effort has been devoted to identifying these vulnerabilities in practical QKD devices and developing corresponding attack strategies. Among all components, single-photon detectors were historically the most vulnerable, and their imperfections have been heavily exploited by various quantum hacking attacks~\cite{Lydersen2010,7571108,PhysRevLett.117.250505,Chistiakov:19,PhysRevA.99.062315,PhysRevA.74.022313,Lydersen:10,Wiechers_2011,Weier_2011,PhysRevA.84.032320,Lydersen_2011,Sauge:11,PhysRevA.91.062301}. Fortunately, the measurement-device-independent (MDI) QKD protocol guarantees that imperfections in measurement devices—including single-photon detectors—do not compromise the overall security of the QKD system~\cite{PhysRevLett.108.130503}. As a result, the security and performance of transmitter modules have become increasingly critical for practical QKD implementations.

In recent years, numerous attacks targeting QKD transmitters have been proposed, including the Trojan-horse attack (THA)~\cite{doi:10.1080/09500340108240904,PhysRevA.73.022320,Jain_2014,PhysRevX.5.031030,Tamaki_2016}, the laser-seeding attack~\cite{PhysRevA.92.022304,PhysRevApplied.12.064043,PhysRevApplied.20.044005}, and the induced-photorefraction attack~\cite{PhysRevApplied.19.054052,Lu:23,PhysRevApplied.20.044013}. In these attacks, Eve injects external light to obtain encoding information through back-reflected photons, manipulate the behavior of the laser source, or alter the operational characteristics of phase and intensity modulators. To mitigate such threats, researchers have deployed essential passive optical components—such as attenuators, isolators, and wavelength-division multiplexers (WDMs)—to limit the amount of injected light entering the transmitter, preventing Eve from extracting secret information~\cite{PhysRevX.5.031030,PhysRevApplied.13.034017}. However, recent studies have revealed that the attenuation of optical attenuators and the isolation of optical isolators can degrade under high-power optical illumination~\cite{PhysRevApplied.13.034017,PRXQuantum.3.040307}, potentially allowing Eve to revive light-injection attacks and compromise the security of the QKD system.

The WDM, another widely used passive component, play an important role in a practical QKD system by multiplexing quantum-state pulses and synchronization pulses before transmission through the quantum channel~\cite{Chen2021, Clivati2022, Zhou2023, PhysRevLett.130.210801, Pittaluga2025}. Since the WDM is directly connected to the quantum channel, it is the first optical device encountered by Eve when launching laser-injection attacks against the QKD transmitter. Therefore, understanding the behavior of WDMs under such attacks is essential for comprehensively assessing the practical security of QKD systems. Nevertheless, the potential security loopholes introduced by WDMs have remained largely unexplored.

In this work, we conduct a detailed investigation into the behavior of WDMs under laser-damage attacks. Our experimental results show that when the injected high-power laser exceeds a certain threshold, its wavelength can shift after passing through specific WDM samples, leading to the appearance of additional spectral peaks at the output. This behavior potentially opens opportunities for Eve to exploit. Through further theoretical analysis, we demonstrate that this spectral side channel can reduce the maximum secure transmission distance to below $66.9\%$ of its original value. By combining experimental observations with theoretical modeling, this work uncovers the security implications introduced by WDMs in practical QKD systems and provides insights for designing more secure QKD transmitters.

The remainder of this paper is organized as follows. Section~\ref{sec:Method} describes the experimental setup and measurement procedures for the laser-damage attack on WDMs. The experimental results are presented in Sec.~\ref{sec:Result}. Section~\ref{sec:Analysis} analyzes the impact of the revealed spectral side channel on QKD security, using the THA as a representative example of light-injection attacks. Finally, Sec.~\ref{sec:Conclusion} discusses potential countermeasures and concludes the paper.

\section{Testing method} \label{sec:Method}

The WDM sample under test features three ports labeled COM, Ref, and Pass. Specifically, the Pass port transmits light within its operating wavelength range (center wavelength $\pm$ bandwidth) while exhibiting high isolation for other wavelengths, whereas the Ref port exhibits the opposite behavior. In the QKD setup, Alice’s quantum-state pulses enters through the Pass port, while the synchronization light enters through the Ref port. The WDM multiplexes these two types of signals and passes them to the quantum channel through the COM port. To examine the optical behavior of the WDM under high-power illumination, we inject a high-power laser into the COM port to simulate Eve’s light entering from the quantum channel.

Depending on the direction of light propagation through the WDM, the experiment is divided into two scenarios. In the first scenario, we analyze how the characteristics of the injected high-power laser change after passing through the WDM and reaching the inside of transmitter. This scenario corresponds to Eve injecting light from the quantum channel and is discussed in detail in the main text. In contrast, the second scenario investigates how the characteristics of light emitted by the transmitter are altered after passing through a WDM that has been exposed to high-power laser illumination. As this case serves as a complementary analysis, it is presented in Appendix~\ref{Details of TLA test} for clarity and conciseness.

\subsection{Experimental setup}

\begin{figure}
\includegraphics[]{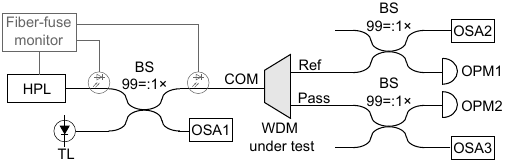}
\caption{Schematic diagram of the experimental setup, with the WDM as a replaceable sample under test. All fibers are standard single mode. OSA, optical spectrum analyzer; OPM, optical power meter; COM, common port; Pass, pass port; Ref, reflect port; TL, tunable laser; HPL, high-power laser; BS, beam splitter. The coupling ratio of the BS denoted $99=:1\times$ means that $99\%$ of light passes to the port horizontally opposite in the graphical symbol of the BS, while $1\%$ of light is coupled across to the other port. }
\label{fig:mini_ILA_Final} 
\end{figure}

The experimental setup is shown in Fig.~\ref{fig:mini_ILA_Final}. In this setup, a high-power laser (HPL) emitting \SI{1550.7}{\nano\meter} continuous-wave light is connected to the COM port of the WDM via a 99:1 beam splitter (BS), simulating Eve's injection of high-power laser light from the quantum channel into the transmitter. Optical spectrum analyzer 1 (OSA1), OSA2, and OSA3 are connected to their respective BSs to measure the spectral characteristics. OSA1 monitors the input spectrum at the COM port, while OSA2 and OSA3 record the output spectra at the Ref and Pass ports, respectively. Optical power meter 1 (OPM1) and OPM2 are used to measure the optical power at the Ref and Pass ports. A tunable laser (TL), which provides a constant-power laser with adjustable wavelength, is also connected to the COM port of the WDM via a 99:1 BS to characterize the initial spectral response of the WDM sample. Additionally, to prevent equipment damage caused by fiber fuse events, a fiber fuse monitor is applied in the setup. 

Regarding the fiber fuse monitor used in our experiments, it is designed to receive light from free space rather than being directly spliced into the optical path. In the event of a fiber fuse, leaked light is captured by the monitor, which then triggers a sensor to cut off the light source, thereby protecting both personnel and equipment. Since the monitor is not directly integrated into the transmission line, it introduces no additional attenuation, spectral distortion, or power fluctuations, and thus does not influence the observed WDM behavior.

\subsection{Test procedure}

Before conducting the main experiment, we first characterize the stability of the HPL, which is the core device used in the laser-damage attack. The HPL provides continuous-wave (c.w.) light with a maximum output power of \SI{10}{W}, and its stability test is described in detail in Appendix~\ref{Details of high-power laser stability test}. After confirming stable output, we proceed with the subsequent measurements. Prior to testing the WDM samples, we verified that the remainder of the setup—excluding the WDMs themselves—exhibits no significant changes under high-power illumination, even when operating at the maximum output power of the HPL. This ensures that any observed variations in subsequent experiments can be attributed solely to the WDM sample under test.


A ``successfully hacked case’’ is defined as any instance in which the spectrum of the light exhibits significant differences before and after passing through the WDM. Before enabling high-power illumination, we switch off the HPL and turn on the TL to measure the initial spectral characteristics of the WDM sample. The tuning range of the TL is determined by the central wavelength and channel passband of each WDM sample. To obtain the complete initial spectrum, the TL must sweep across the full operating wavelength range of the WDM. For example, for a WDM with the central wavelength of \SI{1550.12}{nm} and the passband of ±\SI{0.11}{nm}, the TL is swept from \SI{1548}{nm} to \SI{1552}{nm} in \SI{20}{pm} increments, illuminating the device for 3 seconds at each step. During this sweep, OSA2 and OSA3 operate in hold-max mode, with a scan speed of \SI{200}{nm/s} and a sampling resolution of \SI{2}{pm}, thereby recording the maximum response at each wavelength and yielding the initial spectral profile of the WDM.

Afterward, the TL is turned off and the HPL is turned on. OSA2 and OSA3 are switched to live mode to record the real-time spectrum during the laser-damage attack. The HPL power starts at \SI{1}{W} and increases in increments of approximately \SI{1}{W}, with continuous illumination of up to 10 minutes at each level. If the spectrum observed on OSA2 or OSA3 satisfies the previously defined criteria for a successfully hacked case, the experiment is immediately terminated. If the HPL reaches its maximum output of 10 W without causing notable spectral changes, the experiment is also concluded.
        
\section{Testing Results} \label{sec:Result}
            
\subsection{Spectral feature under test}
\label{ILA}
            
Three types of thin-film WDMs (WDM1–3) were tested in the experiment, and two independent samples were evaluated for each type to ensure experimental reliability. The detailed parameters of all WDM samples are listed in Table~\ref{tab: receiver end WDM parameter}. The experimental procedures followed the methods described in Sec.~\ref{sec:Method}. Among the three types of WDMs, only WDM1 exhibited successfully hacked cases as defined in Sec.~\ref{sec:Method}. Furthermore, for each WDM type, the two independent samples produced consistent test results. In the following, we present the detailed experimental results of sample~1 as a representative of the WDM1 type. Since sample~2 showed similar behavior, its results are summarized only briefly. Samples~3 and~5 are then used as the representatives of WDM2 and WDM3, respectively.

\begin{table*}
\vspace{-0.7em} 
\caption{\label{tab: receiver end WDM parameter}%
The detailed parameters of the WDM samples to be tested in the experiment}
\begin{ruledtabular}
\begin{tabular}[t]{ccccccc}
Type & Sample & Multiplexing Technology & Central Wavelength(nm) & Bandwidth(GHz) & Insertion Loss(dB) & Isolation(dB) \\
\hline
WDM1 & 1,2 & DWDM & 1550.12 &  $100$ & $ < 0.5 $ & $>30$ \\
WDM2 & 3,4 & DWDM+CWDM & 1550.12 &  $100$ & $ < 0.5 $ & $>30$ \\
WDM3 & 5,6 & DWDM+CWDM & 1549.32 &  $100$ & $ < 0.5 $ & $>30$ \\
\end{tabular}
\end{ruledtabular}
\end{table*}        
        
\begin{figure}
\includegraphics[]{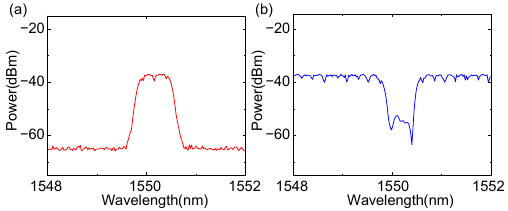}
\caption{The initial spectral of sample 1. (a) Output spectrum at the Pass port. (b) Output spectrum at the Ref port. }
\label{fig:DWDM1_ini} 
\end{figure}

Figures~\ref{fig:DWDM1_ini}(a) and (b) show the spectra at the Pass and Ref ports of sample~1 in the absence of the laser-damage attack. The initial spectrum confirms that both the central wavelength and the channel passband of the WDM are consistent with the specifications provided in the manufacturer's datasheet.

\begin{figure}
\includegraphics[]{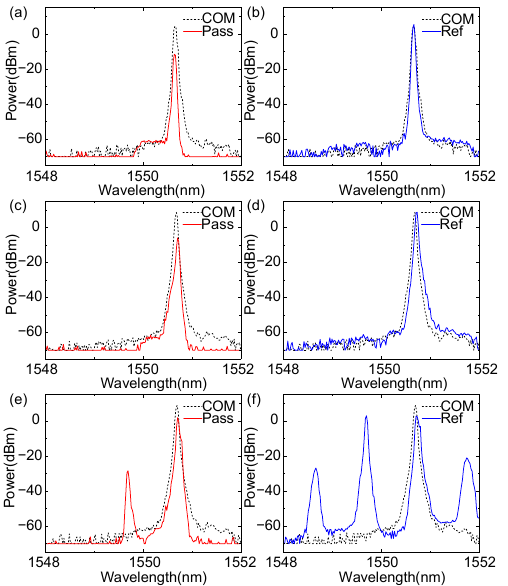}
\caption{The experimental spectral results for sample 1. The black dashed line represents the input spectrum at the COM port, measured by OSA1. The red and blue solid lines represent the output spectra at the Pass and Ref ports, measured by OSA3 and OSA2, respectively. (a) and (b) show the experimental spectral results when the HPL power is \SI{1}{W}; (c) and (d) correspond to the results at \SI{4}{W}; and (e) and (f) present the results at \SI{5}{W}.}
\label{fig:DWDM1_attack} 
\end{figure}

Figure~\ref{fig:DWDM1_attack} shows the spectral response of the WDM under the laser-damage attack. At the HPL power of \SI{1}{W}, as shown in Fig.~\ref{fig:DWDM1_attack}(a) and (b), the Pass-port output exhibits only a reduction in optical power compared to the input at the COM port. The spectral shape at the Ref port remains nearly unchanged and closely matches the input spectrum. The output wavelengths at both ports coincide with the high-power laser wavelength of approximately \SI{1550.7}{nm}. Although the Pass port is not designed to transmit light at this wavelength, a small fraction of the incident light still leaks through due to the limited isolation of the WDM, which is expected. When the HPL power is increased to \SI{4}{W}, the output spectra remain qualitatively similar to those observed at \SI{1}{W}, as shown in Fig.~\ref{fig:DWDM1_attack}(c) and (d). The main differences are an increase in output power and a broader spectral shape, consistent with the spectral characteristics of the HPL at this power level.

A striking change occurs when the HPL power reaches \SI{5}{W}. As shown by the dashed curve in Fig.~\ref{fig:DWDM1_attack}(e), the spectrum at the COM port still exhibits a single peak at \SI{1550.7}{nm}, indicating that the wavelength of the injected laser remains stable. In contrast, the spectra at both the Pass and Ref ports undergo significant changes, with additional peaks emerging at wavelengths distinct from the original \SI{1550.7}{\nano\meter} source. Specifically, for the Pass port output as illustrated by the solid curve in Fig.~\ref{fig:DWDM1_attack}(e), a new peak appears near \SI{1549.7}{\nano\meter} alongside the original peak at \SI{1550.7}{\nano\meter}. Regarding the Ref port output shown in Fig.~\ref{fig:DWDM1_attack}(f), the spectrum exhibits multiple peaks; those at \SI{1549.7}{\nano\meter} and \SI{1550.7}{\nano\meter} are prominent and comparable in height, while the side peaks at \SI{1548.7}{\nano\meter} and \SI{1551.7}{\nano\meter} are lower yet symmetric in magnitude, creating an overall profile that is approximately axisymmetric around \SI{1550.2}{\nano\meter}.

However, the output spectrum from the Pass port in Fig.~\ref{fig:DWDM1_attack}(e) displays a marked asymmetry, with a power difference exceeding \SI{20}{\decibel} between the \SI{1549.7}{\nano\meter} and \SI{1550.7}{\nano\meter} peaks. This deviation occurs because neither wavelength falls within the primary transmission band of the Pass port, meaning both should theoretically be suppressed. Nevertheless, since the HPL operates specifically at \SI{1550.7}{\nano\meter}, the combination of this high incident power and the finite isolation capability of the Pass port results in significant residual leakage. Crucially, the leakage originating from the high-power \SI{1550.7}{\nano\meter} component is substantially stronger than that from the \SI{1549.7}{\nano\meter} component, leading to the observed asymmetry where the \SI{1550.7}{\nano\meter} peak is markedly higher.

\begin{figure}
\includegraphics[]{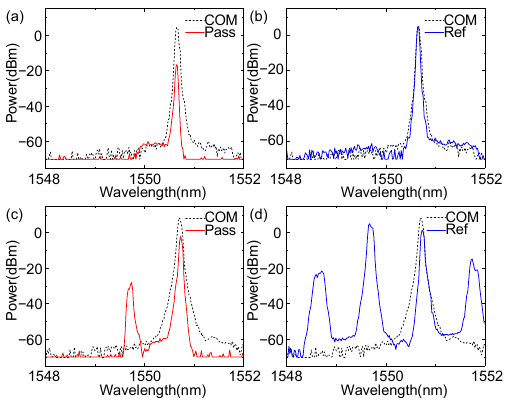}
\caption{The experimental spectral results for sample 2. The black dashed line represents the input spectrum at the COM port, measured by OSA1. The red and blue solid lines represent the output spectra at the Pass and Ref ports, measured by OSA3 and OSA2, respectively. (a) and (b) show the experimental spectral results when the HPL power is \SI{1}{W}, while (c) and (d) present the results at an HPL power of \SI{6}{W}.}
\label{fig:DWDM7_attack} 
\end{figure}
\begin{figure}
\includegraphics[]{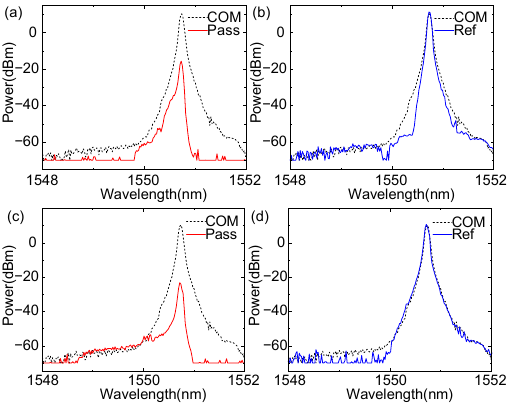}
\caption{The experimental spectral results for samples 3 and 5. The black dashed line represents the input spectrum at the COM port, measured by OSA1. The red and blue solid lines represent the output spectra at the Pass and Ref ports, measured by OSA3 and OSA2, respectively. (a) and (b) show the experimental spectral results of sample 3 under the HPL power of \SI{10}{W}, while (c) and (d) present the corresponding results for sample 5 at the same power level.}
\label{fig:sample3and5} 
\end{figure}

For samples 2, 3, and 5, the initial spectra are not illustrated here because all of them are consistent with the specifications stated in their datasheets. Figure~\ref{fig:DWDM7_attack} shows the input and output spectra of sample 2 under HPL powers of \SI{1}{W} and \SI{6}{W}. When the HPL power reaches \SI{6}{W}, sample 2 exhibits the same phenomenon observed in sample 1 --- multiple spectral peaks appear in the outputs from both the Pass and Ref ports, as illustrated in Fig.~\ref{fig:DWDM7_attack}(c) and (d). For samples 3 and 5, measurements were performed with HPL powers ranging from \SI{1}{W} up to the maximum of \SI{10}{W}. In these tests, no significant multi-peak features were observed, and therefore we present only the results at the maximum HPL power of \SI{10}{W}. Figure~\ref{fig:sample3and5} shows the input and output spectra of samples 3 and 5 under this condition. Compared with the input spectrum at the COM port, the output at the Pass port only shows a reduction in power, while the output at the Ref port remains essentially unchanged. These results indicate that both samples retain their normal spectral characteristics under illumination by the \SI{10}{W} high-power laser, with no noticeable spectral distortions or abnormal features.

Our experimental results show that the samples of WDM1 exhibit significant spectral changes when subjected to the laser-damage attack. In particular, the wavelength of the transmitted high-power laser shifts, resulting in a multi-peak structure in the output spectrum. In contrast, the samples of WDM2 and WDM3 do not display this behavior. Even when the HPL power reaches \SI{10}{W}, their output spectra remain unchanged and no multi-peak features are observed.


\subsection{Possible mechanism}
\label{Analysis_test}

The experiment on WDM1 shows that when the injected high-power laser reaches a sufficiently high level, its wavelength changes after passing through the WDM, producing a multi-peak structure in the spectra observed at both the Pass and Ref ports. The wavelengths of these newly appeared peaks differ from that of the incident high-power laser, indicating that the WDM exhibits nonlinearly distorted under high optical power, effectively redistributing the input energy into multiple wavelength components. In addition, the power of these components depends on wavelength, i.e., the peak closest to the original wavelength has the highest power, while the others are weaker and symmetrically distributed around it. Because detailed information about the internal structure and material properties of WDM1 is not available from the manufacturer, we cannot provide a fully conclusive explanation for this phenomenon. Nevertheless, several key features can be identified from the experimental observations. 

First, the newly generated spectral components are symmetrically distributed around the original wavelength, which is consistent with nonlinear parametric processes such as four-wave mixing (FWM). Second, the spectral peaks exhibit an approximately uniform spacing of about \SI{1}{nm}, corresponding to roughly \SI{125}{GHz} in the frequency domain. Notably, the WDM device has a nominal channel bandwidth of approximately \SI{100}{GHz}, which is of the same order of magnitude as the observed spacing. This correspondence indicates that the characteristic frequency scale of the device may influence the spectral positions of the generated components, although this interpretation remains a hypothesis in the absence of detailed device information.

In addition, modulation instability (MI) may also contribute to the observed spectral features. Near the edge of the passband, optical filters can exhibit strong dispersion, which, combined with high optical power ($\sim$5~W), may give rise to MI. A characteristic feature of MI is the preferential amplification of frequency components at specific offsets determined by the interplay between nonlinearity and dispersion. Subsequent nonlinear interactions, such as cascaded FWM, may further generate additional spectral components, leading to a comb-like multi-peak structure.

Overall, the approximately 1~nm spacing is consistent, at the order-of-magnitude level, with both the intrinsic bandwidth of the WDM device and the characteristic frequency offsets associated with nonlinear effects. This agreement provides a plausible, though not yet conclusive, explanation for the generation of the symmetric multi-peak spectrum observed in the experiment.

\section{Security Analysis} \label{sec:Analysis}

The experimental results presented in Sec.~\ref{sec:Result} demonstrate that laser-damage attack on the WDM may introduce spectral side channel that could be exploited by Eve to implement light-injection attacks. In this section, we analyze the additional security threats this spectral side channel may introduce to practical QKD systems, using the THA as a case study.

In THA, Eve sends Trojan-horse photons along the quantum channel to the transmitter. A portion of Trojan-horse photons is encoded with key information and then reflected back to Eve. Eve extracts encoding information from the Trojan-horse photons reflected back to her, thereby compromising the security of the QKD system~\cite{doi:10.1080/09500340108240904,PhysRevA.73.022320,Jain_2014,PhysRevX.5.031030,Tamaki_2016}. As noted in Ref~\cite{PhysRevX.5.031030}, the amount of information that Eve can extract from the source is directly influenced by the number of Trojan-horse photons reflected back to her. This number depends on three main factors—the intensity of the Trojan-horse light injected into the transmitter, the isolation provided within the transmitter, and the internal reflectivity of its components.
             
The Trojan-horse light injected by Eve from the quantum channel into the transmitter, after passing through the WDM and heading towards the encoding module, can be observed from our experimental results as shown in Fig.~\ref{fig:DWDM1_attack}. To evaluate the transmitter's isolation against Trojan-horse light, we adopt the methodology outlined in Ref.~\cite{PhysRevApplied.22.044076} for measuring the isolation of attenuators and isolators. Detailed procedures are provided in Appendix~\ref{appendix: isolation}. Furthermore, we pessimistically assume an internal reflection efficiency of $1$ at the transmitter, representing the most favorable scenario for Eve.

We utilize the theoretical framework for security analysis outlined in Refs.~\cite{PhysRevX.5.031030,Tamaki_2016} to quantitatively analyze the impact of THA on the security of QKD systems. Specifically, in this work, we analyzed the impact of the combined laser-damage and THA attacks on the security key rate of weak + vacuum decoy-state BB84 QKD protocol~\cite{PhysRevA.72.012326}. Based on Ref.~\cite{1365172}, the key generation rate with weak coherent source (WCS) can be expressed as 
\begin{align}
R \geq q \{ -Q_{\mu} H_2(E_{\mu}) f(E_{\mu}) + P_{1} Y_{1}^{\mu} [1 - H_2(e^{\mu}_{1})] \}.
\label{eq:GLLP}
\end{align}
Here, $q$ depends on the specific protocol, for instance, $q = 0.5$ for the typical BB84 protocol; $\mu$ represents the average photon number of signal state; $Q_{\mu}$ and $E_{\mu}$ respectively represent the total gain and error rate of the signal state; $Y_{1}^{\mu}$ and $e^{\mu}_{1}$ are the yield and error rate of single-photon pulses; $P_{1}$ is the probability of single-photon pulses; $f(x)$ is the bidirectional error correction efficiency, and $H_2(x) = -x \log_2(x) - (1 - x) \log_2(1 - x)$ is the binary Shannon information entropy. 
            
In Eq.(\ref{eq:GLLP}), $Q_{\mu}$ and $E_{\mu}$ can be calculated by communicating parties. Therefore, to assess the lower bound of the key rate, we should estimate the lower bound of $Y_{1}^{\mu}$ and the upper bound of $e^{\mu}_{1}$. Reference~\cite{PhysRevA.72.012326} provides the expressions for $Y_{1}^{\mu}$ and $e^{\mu}_{1,\overline{THA}}$ in the absence of THA as follows.
\begin{gather}        
Y_{1}^{\mu} \geq Y_1^{\mu,L} = \frac{\mu}{\mu\nu - \nu^2} \left( Q_{\nu}e^{\nu} - Q_{\mu}e^{\mu}\nu^2 - \frac{\mu^2 - \nu^2}{\mu^2} Y_0 \right),  \notag \\
e^{\mu}_{1,\overline{THA}} \leq  \frac{E_{\nu}Q_{\nu}e^{\nu} - e_0Y_0}{Y_1^{\mu,L} \nu}.
\label{eq:Decoy_Ye}
\end{gather}
Here, $\mu$ ($\nu$) represents the average photon number of signal state (decoy state); $Q_{\mu}$ and $E_{\mu}$ ($Q_{\nu}$ and $E_{\nu}$) respectively represent the total gain and error rate of signal state (decoy state); $Y_0$($e_0$) is dark count rate (dark error rate).

However, Eve's THA creates a discrepancy between phase errors and bit errors. Reference~\cite{PhysRevX.5.031030}, building on the Gottesman-Lo-Lütkenhaus-Preskill framework~\cite{1365172, lo2007security}, derived the specific expression for parameter $e^{\mu}_{1}$ under the influence of THA as follows
\begin{align}
e^{\mu}_{1} &= {e^{\mu}_{1,\overline{THA}}} + 4\Delta'(1 - \Delta')(1 - 2{e^{\mu}_{1,\overline{THA}}}) \nonumber \\
&\quad + 4(1 - 2\Delta')\sqrt{\Delta'(1 - \Delta'){e^{\mu}_{1,\overline{THA}}}(1 - {e^{\mu}_{1,\overline{THA}}})} ,\nonumber \\
\Delta' &= \frac{\Delta}{Y_{1}^{\mu}},\nonumber \\
\Delta &= \frac{1}{2} \left[1 - \exp(-\mu_{\text{out}}) \cos(\mu_{\text{out}})\right].
\label{eq:PRX_e}
\end{align}
Here, $\mu_{\text{out}}$ represent the intensity of the Trojan-horse light, and $\Delta$ denotes the probability that Alice measures the quantum coin in the $X$-basis and obtains the state $\ket{1}_X$. This concept, central to the GLLP security proof~\cite{1365172}, Koashi’s security proof~\cite{Koashi_2009}, and loss-tolerant protocol~\cite{PhysRevA.90.052314}, is utilized within a virtual protocol to evaluate the upper bound on the information potentially leaked to Eve. Further details regarding $\Delta$ and $\Delta'$ can be found in Refs.~ \cite{1365172,PhysRevA.90.052314,PhysRevX.5.031030}.
            
\begin{figure}
\includegraphics{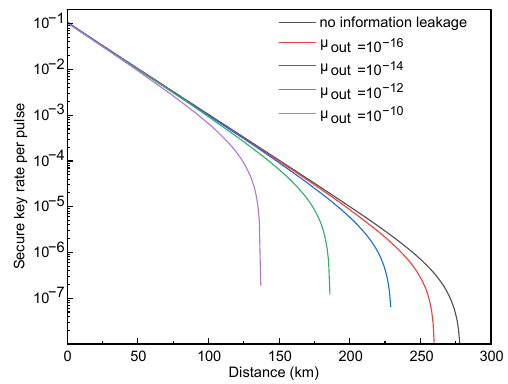}
\caption{Key rate under different intensities of Trojan-horse photons. The main parameters used in the simulation are as follows. Fiber loss coefficient of 0.2 dB/km, Bob’s detector detection efficiency of 42\%, optical error rate of 1\%, dark count probability per gate of $8 \times 10^{-8}$, bidirectional error correction efficiency of 1.16 and dark error rate $e_0$ of 0.5.}
\label{fig:key rate} 
\end{figure}

According to Ref.~\cite{Tamaki_2016}, THA can learn not only phase-encoded information but also intensity-encoded information, making the signal and decoy states distinguishable. It also suggests using trace distance to quantitatively characterize the distinguishability of Trojan-horse photons on signal and decoy states, with the formula being
\begin{align}
\left| {Y_1^{\mu} - Y_1^{\nu}} \right| \leq D_{\mu,\nu}, \nonumber \\
\left| Y_1^{\mu}e^{\mu}_{1} - Y_1^{\nu}e^{\nu}_{1}\right| \leq D_{\mu,\nu}.
\label{eq:Dis_trace}
\end{align}
$D_{\mu,\nu}$ denotes the trace distance between the probability distributions associated with the signal state and the decoy state. Following Ref.~\cite{PhysRevA.98.012330}, this quantity is first evaluated based on the deviation between the quantum states emitted by the source in the presence of source errors and these in the ideal case. Subsequently, one can estimate the lower bound of $Y_{1}^{\mu}$ and the upper bound of $e^{\mu}_{1}$ under the risks of distinguishable decoy state. Subsequently, by integrating Eq.(\ref{eq:GLLP})-(\ref{eq:PRX_e}), one can estimate the key rate under the influence of THA. Figure~\ref{fig:key rate} depicts the curve of the secure key rate under different intensities of Trojan-horse photons.

We integrate the Pass port results of the sample~1 from the experiment (Fig.~\ref{fig:DWDM1_attack}(e)) with the security theoretical model to quantitatively evaluate the maximum key generation distance across different wavelengths. When Trojan-horse photons pass through the WDM, wavelength shifts occur, resulting in significant photon populations at both \SI{1549.7}{nm} and \SI{1550.7}{nm}. The photon count at the original Trojan-horse wavelength (\SI{1550.7}{nm}) is substantially higher than that of the side-channel component (\SI{1549.7}{nm}), implying greater information leakage and a more immediate threat to QKD security, as extensively analyzed in Ref.~\cite{PhysRevX.5.031030}. However, practical countermeasures against Trojan-horse attacks are often designed to suppress photons at the expected attack wavelength. For instance, additional isolators, circulators, or wavelength-selective filters can be employed to increase the isolation at the original Trojan-horse wavelength. Our results show that wavelength-shifted side peaks generated by the WDM may not be effectively suppressed by such wavelength-specific defenses. Consequently, even when the dominant leakage channel is mitigated, Eve may still exploit these side-peak photons to obtain information. Therefore, the subsequent analysis focuses on the impact of these side-peak photons, particularly those at \SI{1549.7}{nm}, on the practical security of QKD.

\begin{figure}
\includegraphics{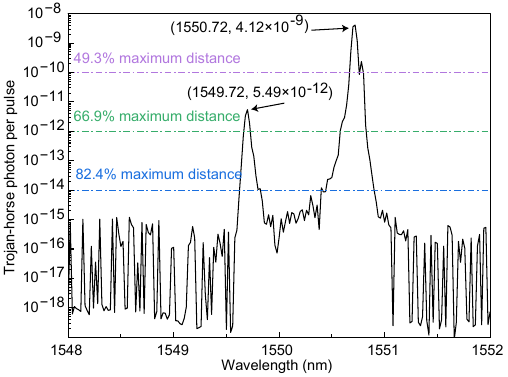}
\caption{The number of Trojan-horse photons in the frequency domain, along with the corresponding maximum distances of key generation.}
\label{fig:error_final} 
\end{figure}

Using the intensity conditions of the Trojan-horse laser assumed in Ref.~\cite{PhysRevX.5.031030}, along with our experimental results on the isolation performance of the isolator, attenuator (detailed in Appendix~\ref{appendix: isolation}), and WDM at different wavelengths, we calculated that the intensity of the Trojan-horse photons is on the order of $10^{-12}$. By incorporating this into the theoretical model, the maximum key generation distance is reduced to 66.9\% of the one without attack. Figure~\ref{fig:error_final} displays the number of Trojan-horse photons in the frequency domain, along with the corresponding maximum distances of secure keys generation for these photons. However, it should be noted that the assumption of unit internal reflection efficiency represents an overly pessimistic, worst-case scenario. In a more practical setting, if the transmitter's internal reflection efficiency is reduced to a realistic value of 1\%, the number of leaked Trojan-horse photons drops significantly to the order of $10^{-14}$. Under these conditions, the secure transmission distance recovers to 82.4\% of the original value, marking a substantial improvement compared to the 66.9\% limit imposed by the worst-case assumption. Our analysis indicates that the discovered spectral side channel could potentially reduce the maximum distance of secure keys generation, representing a significant security risk.

\section{Conclusion} \label{sec:Conclusion}

In this paper, we experimentally investigate the effects of laser-damage attack on WDMs under high-power laser illumination and identify a previously unreported spectral side channel in certain WDM samples. Our theoretical analysis further reveals that this side channel introduces potential security risks to QKD systems. Specifically, when a laser-damage attack is launched at the transmitter, the WDM—being the first optical component exposed to the injected light—can cause part of the high-power laser to undergo wavelength changes as it passes through, resulting in multiple peaks in the output spectrum. Using the Trojan-horse attack as a representative example of light-injection attacks, we analyze the impact of this spectral side channel on a decoy-state BB84 QKD system. Our results show that this side channel can reduce the secure key distribution distance to approximately $66.9\%$ of its original value.

Our experimental results and theoretical analysis demonstrate that the spectral side channel constitutes a significant security vulnerability. A potential countermeasure is to place an isolator at the outermost output of the transmitter, directly interfacing with the quantum channel. This configuration can block a substantial portion of Eve's injected high-power light and thereby prevent the WDM from being directly exposed. It is worth noting that Ref.~\cite{PRXQuantum.3.040307} reports that isolators themselves may also be affected by laser-damage attacks, leading to changes in their characteristics, and suggests that incorporating additional isolators can protect the remaining components of the QKD source against light-injection attacks. Therefore, inserting multiple optical isolators between the WDM and the quantum channel can effectively mitigate device damage and enhance the practical security of QKD systems. Beyond this countermeasure, we suggest that manufacturers incorporate high-power tolerance testing into quality control procedures, where optical components are evaluated under extreme operating conditions to identify potential vulnerabilities. In addition, employing alternative types of WDM filters or monitoring the input optical power also provides a simple yet effective protection strategy. This study offers valuable guidance for designing and implementing more robust QKD systems. Future work could focus on clarifying the mechanism behind the multi-peak phenomenon in WDMs under the laser-damage attack.

\begin{acknowledgments}
{\em Funding:} Quantum Science and Technology-National Science and
Technology Major Project (2021ZD0300704), National Natural Science Foundation of China (62371459), and Hunan Provincial Natural Science Foundation (2026JJ20059). V.M.\ acknowledges funding from the Galician Regional Government (consolidation of research units: atlanTTic and own funding through the ``Planes Complementarios de I+D+I con las Comunidades Aut{\' o}nomas'' in Quantum Communication), MICIN with funding from the European Union NextGenerationEU (PRTR-C17.I1), and the ``Hub Nacional de Excelencia en Comunicaciones Cu{\' a}nticas'' funded by the Spanish Ministry for Digital Transformation and the Public Service and the European Union NextGenerationEU.
\end{acknowledgments}

\begin{appendix}

\section{Experimental details of light emitted from the transmitter under the laser-damage attack on the WDM} \label{Details of TLA test}

In the main text, we rigorously investigated how high-power laser light injected from the quantum channel is altered after passing through the WDM located at the transmitter. To ensure a comprehensive and systematic assessment, this appendix presents a complementary study that specifically examines whether the signal light emitted by the transmitter undergoes any modification after propagating through a WDM subjected to the laser-damage attack. These two core scenarios are mutually reinforcing and together provide a definitive evaluation of the impact of laser-damage attack on the practical security of QKD systems employing WDM components. Furthermore, we conducted an additional supplementary broadband spectral test solely to explore the response of the irradiated WDM to off-center wavelengths; this specific investigation serves only to characterize the filter's behavior across a broader spectrum and does not replace the rigorous validation provided by our primary experiments on the consistent signal path.


\subsection{Setup and test procedure}
\begin{figure}
\includegraphics{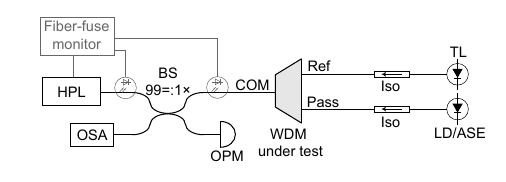}
\caption{Schematic diagram of the experimental setup, with the WDM as a replaceable sample under test. All fibers are standard single mode. OSA, optical spectrum analyzer; OPM, optical power meter; COM, common port; Pass, pass port; Ref, reflect port; TL, tunable laser; LD, laser diode; ASE, amplified spontaneous emission; HPL, high-power laser; Iso, isolator; BS, beam splitter. The coupling ratio of the BS denoted $99=:1\times$ means that $99\%$ of light passes to the port horizontally opposite in the graphical symbol of the BS, while $1\%$ of light is coupled across to the other port.}
\label{fig:mini_TLA_Final} 
\end{figure}

The experiment setup is shown in Fig.~\ref{fig:mini_TLA_Final}. It is noteworthy that this experiment considers the practical application scenarios of the QKD system, focusing on the transmittance of WDM for specific wavelengths of light (quantum-state light at \SI{1550}{nm} and synchronization light at \SI{1570}{nm}). Thus, a $1550$-nm wavelength laser beam is emitted from a laser diode (LD) and transmitted to the Pass port of the WDM. Similarly, the TL emits c.w.\ light with the wavelength at \SI{1570}{nm} to simulate Alice's synchronization light. The COM port of the WDM is connected to the HPL through a 99:1 BS, simulating Eve's laser damage attack on the transmitter via the quantum channel. The OPM is used to measure the input power of the HPL injected to through the COM port. The OSA is used to measure the output spectrum at the COM port of the WDM. To prevent laser damage to the LD and TL during the experiment, isolators are placed between these light sources and WDM. Additionally, to mitigate the risk of equipment damage due to optical fiber fuse, a fiber fuse monitor is also applied in the setup. 

In the experiment, the LD and TL in Fig.~\ref{fig:mini_TLA_Final} remain continuously turned on, emitting light with the wavelengths around \SI{1550}{nm} and \SI{1570}{nm}, respectively. The power of the HPL starts at \SI{0}{W}~\footnote{Unlike the experiments presented in the main text, this test starts with an HPL power of \SI{0}{W}. This is because, in this experiment, other light sources such as the LD are also active. Therefore, when the HPL power is \SI{0}{W}, the spectrum of the WDM without the laser-damage attack can still be observed. In contrast, in the main text, only the HPL serves as the light source (with the TL turned off during formal testing). As a result, when the HPL power is \SI{0}{W}, there is no light source in the setup, making it impossible to measure the WDM spectrum. Thus, the tests in the main text begin with the HPL power set to \SI{1}{W}.} and increases by approximately \SI{1}{W} each time, continuously illuminating for up to \SI{10}{minutes}. Once the spectrum observed on the OSA meets the criteria of a successfully hacked case as defined in Sec.~\ref{sec:Method}, the experiment is immediately terminated. If the HPL power increases to the maximum power with no significant change in spectrum, the experiment is also stopped.

To investigate the changes in the optical properties of light at various wavelengths after passing through a laser-damaged WDM, we designed a wide-spectrum testing using the Pass port as an example. The experimental setup is slightly modified from the previously described configuration, as shown in Fig.~\ref{fig:mini_TLA_Final}, with the difference that an amplified spontaneous emission (ASE) source with wavelength range from \SI{1518.6}{nm} to \SI{1569.4}{nm} was used instead of LD connected to the Pass port. Using ASE allows one to assess whether WDM performance under the laser-damage attack varies in the wide range of wavelengths. Moreover, since the focus of this experiment is on light transmitted through the Pass port, the Ref port is left to be unconnected. The wide-spectrum testing follows the same procedure as the one described above, except that the ASE source is activated in place of the LD and TL.

\begin{figure*}
\includegraphics{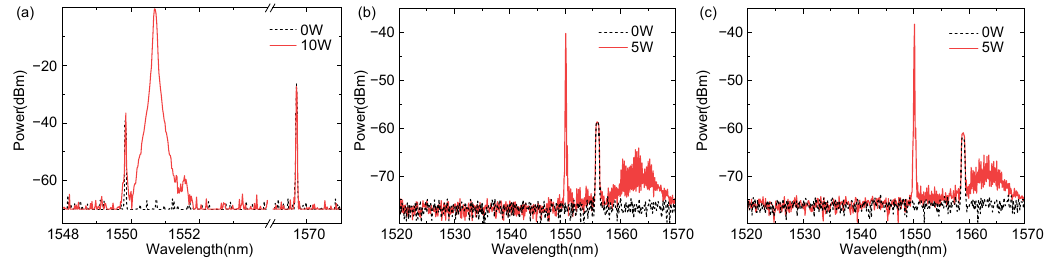}
\caption{(a), (b), and (c) show the output spectra at the COM port for samples 8, 9, and 10, respectively, as measured by the OSA. The black dashed curves correspond to the spectra with the HPL turned off, while the red solid curves represent the spectra with the HPL turned on at its maximum output power, respectively.}
\label{fig:DWDM234} 
\end{figure*}

\begin{figure}
\includegraphics{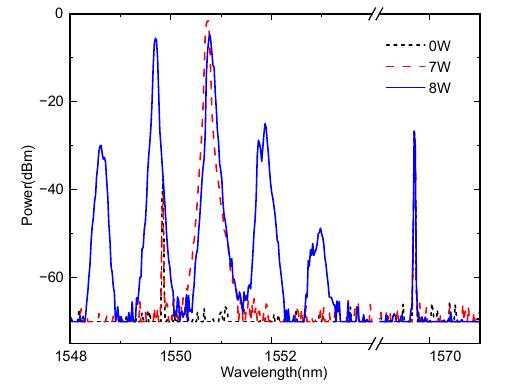}
\caption{The experimental results for sample 7. The black dotted line, red dashed line, and blue solid line represent the spectra at the COM port of the WDM under HPL powers of \SI{0}{W}, \SI{7}{W}, and \SI{8}{W}, respectively.}
\label{fig:res_TLA} 
\end{figure}

\subsection{Result}

In this experiment, four types of thin-film WDM samples (WDM1, WDM2, WDM4, and WDM5) were tested. Table~\ref{tab: source end WDM parameter} presents the detailed parameters of these types of WDM samples. We used the LD with a wavelength near \SI{1550}{nm} and conducted the experiment on WDM1 and WDM2. For WDM4 and WDM5 listed in Table~\ref{tab: source end WDM parameter}, we conducted the wide-spectrum testing, i.e., using ASE to replace the LD in order to achieve a broadband spectrum experiment. Among these, only the sample 7 exhibited the successfully hacked case defined in Sec.~\ref{sec:Method}, while the other three types of WDM samples did not show significant change. 

\begin{table*}
\vspace{-0.7em} 
\caption{\label{tab: source end WDM parameter}%
The detailed parameters of the WDM samples to be tested in experiment}
\begin{ruledtabular}
\begin{tabular}[t]{ccccccc}
Type & Sample & Multiplexing Technology & Central Wavelength(nm) & Bandwidth(GHz) & Insertion Loss(dB) & Isolation(dB) \\
\hline
WDM1 & 7 & DWDM & 1550.12 &  $100$ & $ < 0.5 $ & $>30$ \\
WDM2 & 8 & DWDM+CWDM & 1550.12 &  $100$ & $ < 0.5 $ & $>30$ \\
WDM4 & 9 & DWDM & 1555.74 &  $100$ & $ < 0.5 $ & $>30$ \\
WDM5 & 10 & DWDM & 1558.98 &  $100$ & $ < 0.5 $ & $>30$ \\
\end{tabular}
\end{ruledtabular}
\end{table*}

For the experiment on sample 7, when the HPL power is set to \SI{0}{W}, the COM port spectrum exhibits two peaks near \SI{1550}{nm} and \SI{1570}{nm}, corresponding to the wavelengths of the incident light from the Pass and Ref ports, respectively. We then measured the COM port spectra with the HPL power set to 1, 2, 3, 4, 5, 6, and \SI{7}{W}. The spectrum at \SI{7}{W} is presented as a representative example to illustrate this test procedure. As shown in Fig.~\ref{fig:res_TLA}, when the HPL power reaches \SI{7}{W}, a new peak appears at \SI{1550.7}{nm}, which matches the wavelength of the HPL. The power of this peak increases with the HPL power, indicating that it originates from the HPL and is captured by the OSA due to end-face reflections within the transmitter. Comparing the spectra at \SI{0}{W} and \SI{7}{W}, we observe that the peaks near \SI{1550}{nm} and \SI{1570}{nm} remain nearly unchanged, suggesting that the high-power laser irradiation does not affect the transmission efficiency of the WDM at the Pass and Ref ports.
     
When the HPL power increases to \SI{8}{W}, multiple peaks appear in the 1548--\SI{1553}{nm} range, with intervals of approximately \SI{1}{nm}. Among them, the peaks at \SI{1550}{nm} and \SI{1551}{nm} are the strongest, while the others are weaker and symmetrically distributed. Consistent with the phenomena described in Sec.~\ref{sec:Result}, this spectrum exhibits a distinct multi-peak structure. The power of these additional peaks significantly exceeds that of the LD light transmitted through the Pass port, suggesting they originate from end-face reflections induced by HPL irradiation. We hypothesize that this multi-peak generation results from a nonlinear optical process within the WDM that broadens the single-frequency input into an optical frequency comb.

\begin{figure*}
\includegraphics{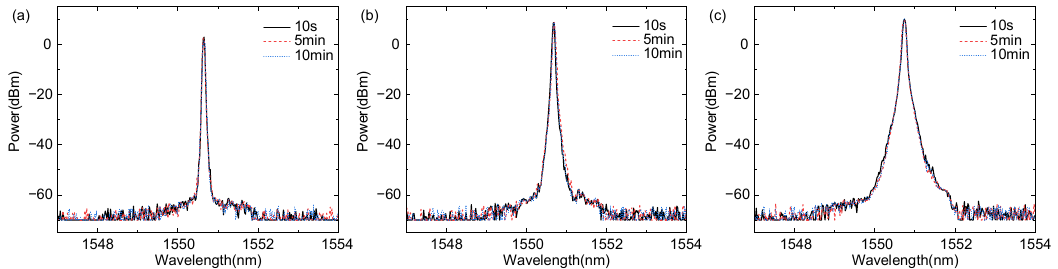}
\caption{Stability test results of the HPL. The black solid line, red dashed line, and blue dotted line represent the spectra when the HPL is turned on for 10 seconds, 5 minutes, and 10 minutes, respectively. (a), (b), and (c) correspond to the spectra under HPL powers of \SI{1}{W}, \SI{5}{W}, and \SI{10}{W}, respectively.}
\label{fig:HPL_test_appendix} 
\end{figure*}

Figure~\ref{fig:DWDM234}(a) presents the test results of the experiment for the sample 8. When the HPL power is set to \SI{0}{W}, the spectrum exhibits two peaks, similar to the case of sample 7. As the HPL power increases to \SI{10}{W}, no spectral changes are observed for the light around \SI{1550}{nm} and \SI{1570}{nm}, and the end-face reflection spectrum matches that of the \SI{10}{W} HPL spectrum shown in Appendix~\ref{Details of high-power laser stability test}, with no multi-peak features. This behavior contrasts with that of sample 7 and indicates that the characteristics of sample 8 remain unaffected under the maximum HPL power used. 

The wide-spectrum measurements of samples 9 and 10 show no significant changes near the signal wavelength relevant to the specific attack investigated in this work. In this experiment, the maximum HPL power used was \SI{5}{W}. As shown in Fig.~\ref{fig:DWDM234}(b) and (c), no distinct peaks are observed near the central wavelength of the WDM, whereas a broad spectral feature appears in the \SIrange{1560}{1570}{\nano\meter} stopband. The purpose of this supplementary experiment was to determine whether the transmitter output undergoes spectral degradation after passing through the HPL-irradiated WDM, with particular emphasis on the integrity of the signal transmitted to the quantum channel. Under the operating conditions considered here, the transmitter signal is centered at \SI{1550}{\nano\meter} within the passband. Therefore, the observed changes in the \SIrange{1560}{1570}{\nano\meter} stopband do not affect the transmitted signal and do not demonstrate the specific attack investigated in the main text. However, these changes may constitute a different spectral side channel under more general attack scenarios, a possibility that is beyond the scope of the present study.


\section{Details of high-power laser stability test} \label{Details of high-power laser stability test}

\begin{figure}
\includegraphics{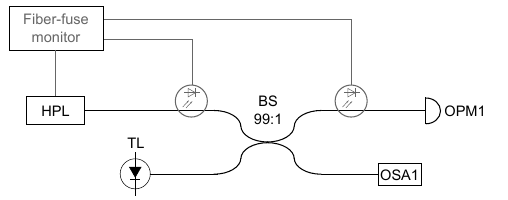}
\caption{Schematic diagram of high-power laser stability test.}
\label{fig:HPL_setup} 
\end{figure}

This section outlines the specifics of the stability test conducted for the HPL, with the test configuration depicted in Fig.~\ref{fig:HPL_setup}. In the experimental setup, the HPL is interfaced with OSA1 and OPM1 via a 99:1 BS. OPM1 serves the purpose of measuring the majority of energy emitted by the HPL, while OSA1 is utilized for real-time monitoring and recording of the live spectrum of the HPL to evaluate its stability. Moreover, integration of a fiber fuse monitor ensures immediate deactivation of the HPL upon detection of optical fiber fuse occurrences, thereby safeguarding the equipment from potential damage. OSA1 remains fixed to specific settings throughout the test. The sensitivity level is set at \SI{-65}{dBm}, the scanning rate is \SI{200}{nm/s}, the scanning mode operates in single-scan mode, and the data recording mode runs in live mode. Each instance of OSA1 scanning results in the recording of the spectrum for subsequent analysis.

In order to evaluate the stability of the HPL, tests were conducted to assess its output fluctuation over both short and long durations. The output power of the HPL starts at \SI{1}{W} and increases incrementally to the maximum value of \SI{10}{W} in a linear manner. At each power level, the laser remains continuously active for 10 minutes, during which OSA1 records the real-time output spectrum of the HPL. Fig.~\ref{fig:HPL_test_appendix}(a), (b), and (c) show the spectral measurements of the HPL at power levels of \SI{1}{W}, \SI{5}{W}, and \SI{10}{W}, respectively. As seen in the figures, the output spectrum of the HPL remains essentially unchanged over time—at 10 seconds, 5 minutes, and 10 minutes after activation. This indicates that the HPL used in the experiment is highly stable and meets the requirements for conducting the laser-damage attack experiments.

\section{Transmitter isolation characterization} 
\label{appendix: isolation}

\begin{figure}
\includegraphics{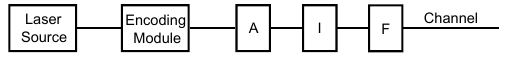}
\caption{Schematic diagram of the internal structure of the transmitter. Various passive optical components are inserted between the encoding device and the quantum channel to enhance isolation, preventing light injection from the channel into the transmitter, which could otherwise compromise the practical security of the QKD system. A: Attenuator; I: Isolator; F: Filter.}
\label{fig:transmitter} 
\end{figure}

\begin{figure}
\includegraphics{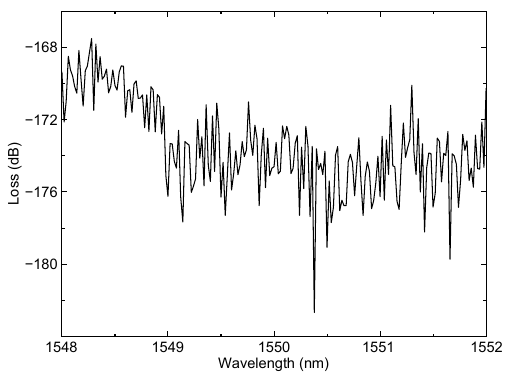}
\caption{The total isolation provided by the attenuator and isolators at different wavelengths.}
\label{fig:loss} 
\end{figure}

Inside the transmitter, attenuators, isolators, and filters are typically configured to increase the isolation from the channel, thereby preventing injected light from the quantum channel from affecting the transmitter, as shown in Fig.~\ref{fig:transmitter}. To evaluate the transmitter's isolation against Trojan-horse light, we follow the measurement method described in Ref.~\cite{PhysRevApplied.22.044076}. Specifically, we use a supercontinuum laser source (NKT Photonics SuperK Fianium FIU-15) and an optical spectrum analyzer (Yokogawa AQ6370D) to measure the broadband transmittance of the optical components. The transmittance of a device under test (DUT) is obtained by first recording the spectrum of the light source alone, followed by a second measurement with the DUT inserted. The DUT transmittance is then calculated as the ratio between the spectral power measured with and without the DUT in place. Using this method, we characterize the broadband transmittance of the attenuator and isolators. The total isolation provided by these components at different wavelengths is shown in Fig.~\ref{fig:loss}.

\end{appendix}

\end{document}